\documentclass[ 
twocolumn,
amsmath,amssymb,
aps,
pre,
longbibliography,
nofootinbib
]{revtex4-1}

\usepackage{graphicx}
\usepackage{color}
\usepackage{subfigure}
\usepackage{hyperref}
\definecolor{red}{rgb}{0.75,0,0}
\definecolor{blue}{rgb}{0,0,0.75}
\definecolor{green}{rgb}{0,0.5,0}
\usepackage{amsmath}
\usepackage{amssymb}
\usepackage{lipsum}
\usepackage{bm}
\usepackage{xcolor}
\usepackage{graphicx}
\usepackage{verbatim}
\usepackage[T1]{fontenc}
\usepackage{hyperref}
\usepackage[normalem]{ulem} %For strikethrough text, can be removed for final version
\def\be{\begin{equation}}
\def\ee{\end{equation}}
\def\bea{\begin{eqnarray}}
\def\eea{\end{eqnarray}}

\def\besub{\begin{subequations}}
\def\eesub{\end{subequations}}

\def\bwd{\begin{widetext}}
\def\ewd{\end{widetext}}

\definecolor{ao(english)}{rgb}{0.0, 0.5, 0.0}
\definecolor{armygreen}{rgb}{0.29, 0.33, 0.13}
\definecolor{auburn}{rgb}{0.43, 0.21, 0.1}
\definecolor{brightmaroon}{rgb}{0.76, 0.13, 0.28}
\definecolor{cadmiumred}{rgb}{0.89, 0.0, 0.13}
\definecolor{carnelian}{rgb}{0.7, 0.11, 0.11}
\definecolor{cornellred}{rgb}{0.7, 0.11, 0.11}
\definecolor{crimsonglory}{rgb}{0.75, 0.0, 0.2}
\definecolor{orangeyellow}{rgb}{0.3, 0.2, 0.2}
\definecolor{fluorescentorange}{rgb}{1.0, 0.75, 0.0}
\definecolor{gamboge}{rgb}{0.89, 0.61, 0.06}
\newcommand{\bsf}[1]{\textsf{\textbf{#1}}}

\newcommand{\AETtext}[1]{#1}
\newcommand{\AETtextB}[1]{\textcolor{blue}{#1}}

\newcommand{\AMrev}[1]{#1}
\newcommand{\red}[1]{#1}

\begin{document}

\title{Front speed and pattern selection \\of a propagating chemical front in an active fluid} 
\author{Clara \surname{del Junco}}
\affiliation{Department of Chemistry and the James Franck Institute, University of Chicago}
\affiliation{Wesleyan University Library, Middletown, Connecticut 06457}
\author{Andr\'e Estevez-Torres and Ananyo Maitra}
\affiliation{Sorbonne Université, CNRS, Institut de Biologie Paris-Seine (IBPS), Laboratoire Jean Perrin (LJP), F-75005, Paris}

\begin{abstract}
Spontaneous pattern formation in living systems is driven by reaction-diffusion chemistry and active mechanics. The feedback between chemical and mechanical forces is often essential to robust pattern formation, yet it remains poorly understood in general. In this analytical and numerical paper, we study an experimentally-motivated minimal model of coupling between reaction-diffusion and active matter: a propagating front of an autocatalytic and stress-generating species. In the absence of activity, the front is described by the the well-studied KPP equation. We find that front propagation is maintained even in active systems, with crucial differences: an extensile stress increases the front speed beyond a critical magnitude of the stress, while a contractile stress has no effect on the front speed but can generate a periodic instability in the high-concentration region behind the front. We expect our results to be useful in interpreting pattern formation in active systems with mechano-chemical coupling {\it in vivo} and {\it in vitro}.
\end{abstract}

\maketitle 

\section{Introduction}

The emergence of form in the living embryo involves a series of out-of-equilibrium chemical and physical processes  \cite{Collinet2021}. Among them, two mechanisms stand out: reaction-diffusion (RD)  \cite{Cross2009, CrossHohen, turing1952chemical, Kondo2010, green2015positional}, which creates well-defined concentration patterns, and active matter (AM), mediated by molecules that convert chemical energy into mechanical forces generating hydrodynamic flows  \cite{gross2017, RMP, CurieRev1, CurieRev2, SRrev, TonTuRam, SR_JSTAT, LPD_JSTAT, CatesRev1, CatesRev2}. 
RD dynamics have also been used to model a wide range of situations in biology ranging from spatiotemporal protein dynamics in cells  \cite{Elias} to the migration of a cell monolayer  \cite{Oreffo, Oreffo1}. In these contexts, the species undergoing RD dynamics may regulate or control active forces in the living system. Indeed, the importance of interactions between chemistry and mechanics in modeling living systems has long been recognized~\cite{turing1952chemical}, yet RD and AM systems have generally been examined separately, whether theoretically \cite{turing1952chemical, RMP} or in {\it in vivo} \cite{Kondo2010, Muller2012} and {\it in vitro} \cite{epstein1998introduction, nedelec1997} experiments. \red{More recently, however, mechano-chemical couplings have been investigated theoretically~\cite{Bois2011, kumar2014}, demonstrating the ability of such systems to display novel patterns even in the absence of chemical reactions\cite{Bois2011}. When combined with chemical reactions, the coupling modifies the patterning behavior of RD systems~\cite{Bois2011}, and can also generate spatiotemporal oscillations~\cite{kumar2014}.} Their importance for {\it in vivo} assays has been established \cite{gross2017, Collinet2021}, and chemical reactions have also been implicated in the active context in spontaneously creating stable droplets of a finite size and formation of protocells  \cite{Zwicker}. However, how active flows affect RD dynamics in general remains sparsely understood.

To the best of our knowledge, couplings between RD and AM have not yet been engineered {\it in vitro}. However, recent experimental developments suggest that mechano-chemical self-organization could be investigated in well-controlled {\it in vitro} experiments in the near future \cite{senoussi2021book}. Such {\it in vitro} systems, which we will call reaction-diffusion active matter (RDAM) systems, would be particularly interesting for quantitatively testing theoretical predictions concerning the propagation of a reacting and diffusing activity regulator -- a situation that is likely to be ubiquitous in {\it in vivo} settings.

In this paper, we present an examination of how active flows affect the propagation of a reacting and diffusing chemical species that controls the strength of activity, using a highly simplified model. We describe the simplest, non-trivial RDAM system that could be reasonably assembled {\it in vitro} with current knowledge (Sec.~\ref{sec_exp_syst}). It is an RD system generating a traveling front that is coupled to a compressible active fluid. We model it with a dynamical equation for the concentration of the reactant and a constitutive equation for the velocity field of the active fluid  \cite{Bois2011} (Sec.\ref{sec_model}).
A linear stability analysis of these equations about a homogeneous state shows that there is a spatial instability for a contractile RDAM system (Sec.~\ref{sec_linear_theory}). However, the linear theory falls short of being able to predict the behavior of the system when symmetry is broken by a  traveling front. In particular, a linear theory incorrectly predicts that the front speed is insensitive to the active stresses. Moreover, it does not provide a prediction of how the spatial instability propagates or how its selected wavelength is shifted in the presence of the front. Numerical simulations of the two dynamical equations provide the answers to these questions (Sec.~\ref{sec_num_results}).  The simulations show that, in extensile fluids, fronts propagate faster as the active stress increases and spatial instabilities are never observed. We show that the front speed picked out in simulations is the minimum physically allowed speed, consistent with a marginal stability mechanism that shows up in much simpler systems, but which does not hold in our highly nonlinear model. In turn, contractile fluids generate fronts whose velocity is independent of the active stress, and they display spatial instabilities whose group velocity is not equal to the speed of the front and whose wavenumber is shifted to smaller values compared to the linear prediction. 

\section{In vitro reaction-diffusion active matter \label{sec_exp_syst}}

In this section we will describe an RDAM system that could be assembled {\it in vitro} with the current experimental state of the art. We will take into account known experimental constraints to motivate the theoretical description undertaken in the rest of the paper. Although a wide array of AM experimental systems exist, we will restrain ourselves to molecular active matter because we believe that it has the highest chance to be combined with molecular RD systems. In this category, extracts of cytoskeletal filaments and motors, either actomyosin or kinesin/microtubule gels \cite{Andorfer2019, senoussi2021book}, are particularly suitable. Depending on the situation, these gels are locally contractile or extensile and are composed of polar, nematic or isotropic particles \cite{needleman2017}. A wide variety of dissipative behaviors have been observed in these gels {\it in vitro}, such as aster formation \cite{nedelec1997}, global contractions \cite{bendix2008quantitative, foster2015active}, chaotic flows \cite{sanchez2012}, active nematics \cite{sanchez2012} and active film buckling \cite{senoussi2019}. These systems function in aqueous solution, at neutral pH and room temperature.

 Chemical systems that generate RD patterns have historically relied on strongly acidic reactions \cite{epstein1998introduction} which are incompatible with cytoskeletal extracts. However, in the last 15 years, biochemical RD systems working at neutral pH have been developed  \cite{isalan2005engineering,loose2008spatial,chirieleison2013pattern,padirac2013spatial,tayar2015propagating}. Among them, DNA-only \cite{chirieleison2013pattern} and DNA/enzyme \cite{padirac2013spatial} reaction schemes are particularly interesting because the modularity of the DNA hybridization reaction facilitates the engineering of mechano-chemical couplings with active fluids. In particular, one may envision that the single-stranded (ssDNA) species A produced by these reactions dimerizes DNA-motor conjugates \cite{wollman2014, Tayar2021}, or releases them from a reservoir \cite{Vyborna2021}. Indeed, a number of active fluids become active only in the presence of dimers of motors \cite{nedelec1997, sanchez2012} and thus, with this scheme, active stresses would be generated only in the presence of species A. 
 
 With such a design, we will consider an autocatalytic reaction of ssDNA species A, which has been reported to generate RD fronts propagating at constant velocity in a thin microchannel \cite{zadorin2015} and that is compatible with kinesin-microtubule active matter \AETtextB{ \cite{senoussi2021_RDAMbis}}. In the following, we will suppose that this autocatalytic reaction takes place in a solution that is in contact with a thin film of an active fluid composed of filaments and motor-DNA conjugates. We note that active thin films, as well as motor-DNA conjugates, have been obtained both for myosin \cite{kumar2018,hariadi2015} and kinesin \cite{sanchez2012,martinez2019, furuta2013, wollman2014, sato2017}. Such a thin film of DNA-responsive active fluid could be obtained at a water-oil interface in the presence of a depletion agent, such as polyethyleneglycol, that pushes the filaments against that interface \cite{kumar2018,Andorfer2019}.

\section{Active traveling front model\label{sec_model}}

\begin{figure}
\includegraphics[width = \linewidth]{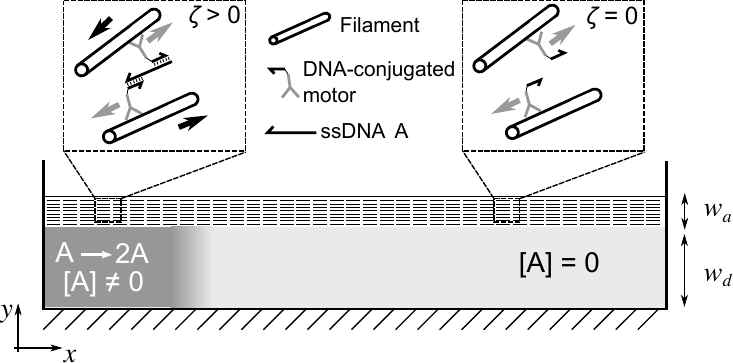}
\caption{Scheme of a plausible experimental reaction-diffusion active matter system that can be described using Eqs. \eqref{eq:cons1mine}-\eqref{eq:fbalmine}. A chamber is filled with a solution containing an RD and an AM subsystem. The RD subsystem is an autocatalytic DNA/enzyme reaction of the form $\textrm{A} \rightarrow 2\textrm{A}$, where A is a ssDNA. The AM subsystem is made of filaments and DNA-conjugated motors. Due to depletion forces, the AM subsystem forms a film of height $w_a$ (segmented pattern) above the RD solution of height $w_d$. A small quantity of A is introduced on the left side of the chamber (dark gray). In the absence of activity (\emph{i.e.} [ATP] = 0 and $\zeta = 0$), A is expected to generate a traveling front of constant velocity\cite{zadorin2015}. In the presence of ATP, A will diffuse on the active region, generating motor dimers that will move along the filaments (gray arrows) generating a flow of filaments (black arrows) and thus induce $\zeta > 0$. At the right side of the front, $[\textrm{A}]=0$, the motors are monomeric and thus they move along the filaments without generating flow $\zeta = 0$.}
\label{fig:scheme}
\end{figure}

In the situation just described, the thin film of active fluid can exchange both momentum and matter (i.e., the filaments can diffuse into the fluid above) with  the solution where DNA/enzyme reactions take place. At a given point in space, as the concentration of A increases in the bulk solution, motor dimers will form in the film and will generate active flows \AETtext{(Fig.~\ref{fig:scheme})}. In order to model this 3-variable (DNA A, motors M and filaments F), biphasic system as a 1-variable monophasic system we will make two sets of assumptions. First, we will suppose that the time-scales of motor-motor dimerization via DNA hybridization, $\tau_{M,A}^+$, of motor-motor dissociation, $\tau_{M,A}^-$, of motor binding to filaments, $\tau_{M,F}^+$, and of motor-filament dissociation, $\tau_{M,F}^-$, are fast compared with the time-scale of DNA autocatalysis, $\tau_A$. Then, one can model such a system with a single \AETtext{average} species C, of concentration $c$, that could be seen as the concentration of motor dimers \emph{bound to filaments} that undergo an autocatalytic reaction (referred to as autocatalytic motors in the following). \AETtext{Note that the effective diffusion coefficient $D$ of this species is an average of the diffusion coefficient of free A, $D_A$, and of A bound to M, $D_M$, weighted by the time spent in each state\cite{zadorin2015}}. Second, we will suppose that the thickness of both the active film, $w_a$, and the DNA/enzyme solution, $w_d$, are small enough such that the corresponding diffusion time $\tau_D = (max(w_a, w_d))^2/2D_A \ll\tau_A$, with $D_A$ the diffusion coefficient of A. As C is advected by active flows in the thin active film, this transport may in turn influence the concentration of C in the bulk by the rapid diffusion of C across both phases. 

To estimate $\tau_{M,A}^+$, we suppose that motor-DNA conjugates dimerize through DNA hybridization with second order kinetics. Thus $\tau_{M,A}^+=1/(\kappa_{M,A}^+ M_0)$, where we take $\kappa_{M,A}^+\sim 10^4-10^6$~M$^{-1}$s$^{-1}$ as a typical value of DNA association rate \cite{estevez2008} and $M_0 \sim 1-100$~nM the typical concentration of kinesin in active matter experiments \cite{senoussi2019}. $\tau_{M,A}^- = 1/\kappa_{M,A}^-$, where the dissociation rate of a DNA double strand $\kappa_{M,A}^-$ is controlled by its length and can vary in the range $\kappa_{M,A}^-\sim10^{-2}-10^{-5}$~s$^{-1}$. $\tau_{M,F}^+=1/\kappa_{M,F}^+$ and we take $\kappa_{M,F}^+ = 5$~s$^{-1}$ from Leduc et al. \cite{leduc2004}. $\tau_{M,A}^- = 1/\kappa_{M,F}^-$, and we calculate $\kappa_{M,F}^- = v/\ell_r$, where $v$ is the velocity of kinesin along microtubules and $\ell_r$ the run length before detachment. Furuta et al. \cite{furuta2013} measured $v$ and $\ell_r$ for DNA-promoted kinesin clusters with different numbers of kinesins and found $v = 800$~nm/min and $\ell_r =2\times10^3$, $4\times10^3$ and $15\times10^3$, respectively for 1-, 2- or 4-kinesin clusters. $\tau_A$ has been measured by Zadorin et al. \cite{zadorin2015} to be in the range $10^2-10^4$~s. Finally $D_A\approx D_M=10^{-10}$~m$^2$s$^{-1}$ and $w_a$ and $w_d$ are in the range $10-100~\mu$m. We thus get $\tau_{M,A}^+ \sim 10-10^5$~s, $\tau_{M,A}^- \sim 10^2-10^5$~s, $\tau_{M,F}^+ \sim 0.2$~s, $\tau_{M,F}^- \sim 2-20$~s, $\tau_A \sim 10^2-10^4$~s and $\tau_D \sim 1-100$~s. As a result, provided that the motor-DNA system is designed such that $M_0$ and $\kappa_{M,A}^-$ are on the upper side of the aforementioned ranges, $\tau_A$ is larger than all the other characteristic times as required by our model.

Beyond the specific and explicit {\it in vitro} scenario described here, the RDAM model we will construct is relevant to a wide range of {\it in vivo} systems. Indeed, our theory describes all active thin films of uniform thickness $h$, where the thickness of the film is maintained by the exchange of material with the surrounding medium  \cite{Salbreaux, Ano_LRO}, and in which activity is controlled by an activity regulator C which in the absence of active flows would have the following RD dynamics: an unstable lower concentration state of the regulator is changed to a higher concentration state at a front moving at a well-defined speed. Stated in this general form, our theory may be relevant for understanding the propagation of a cell layer in a wound-healing assay  \cite{wound-healing}, where C models the cells or for the dynamics of a chemical active stress regulator in the cellular cortex.

Having established the experimental relevance of an RD system coupled to an active flow, we now construct a model for understanding two key features of such a system: i. whether active flows enhance or inhibit the speed of the RD front and ii. whether the profile of the component C behind the front remains homogeneous or forms a patterned state. For this, we consider a system which is perfectly homogeneous in $\hat{y}$ and has a moving front of $C$ along $\hat{x}$ at which a low (zero) concentration state of $C$ changes to a high (non-zero) concentration state. We take $c$ to be the concentration of the species C and $v\equiv v_x\hat{x}$ to be the flow velocity.

With these assumptions, the general equation of motion for the concentration field is
\begin{equation}
\partial_t c = D\partial_{xx} c - \partial_x(vc) + f(c)
\label{eq:cons1mine}
\end{equation}
where $D$ denotes the diffusivity, the second term on the R.H.S. accounts for the advection of the concentration by the flow and $f(c)$ is a generic reaction term with the following properties: i. it destabilizes a steady-state with $c_{ss}=0$ and ii. it saturates at finite value $c_0$. For most of the paper, we will use a simple form for $f(c)$:
\begin{equation}
f(c)=rc(1-c/c_0)
\end{equation}
but we have checked that our results do not depend qualitatively on this particular form. Note that, in the absence of the advection by the velocity field, \eqref{eq:cons1mine} constitutes the well-known equation for a traveling wave introduced independently by Fisher~\cite{Fisher1937} and Kolmogorov, Petrovsky, and Piskunov~\cite{Kolmogorov1991}, which we refer to as the KPP equation~\footnote{We choose not to further promote Ronald Fisher's name due to the racist, ableist, and otherwise supremacist and discriminatory views that he championed in his life and work. He published this equation in the journal {\it Annals of Eugenics} (now {\it Annals of Human Genetics})~\cite{Fisher1937}. We acknowledge that apart from this equation, many of the tools that we use in this paper--even concepts as common as the standard deviation--were developed by Fisher and colleagues of his with similar views, such as Karl Pearson.  We strive and encourage the reader to think critically about the motivations behind their development. For recent discussions in popular fora, see for example~\cite{Levy2019, Evans2020, Gelman2020}. For scholarly studies, see for example Mackenzie's history/social studies of science work~\cite{Mackenzie1978, Mackenzie1981} and a disability and literary studies perspective~\cite{Davis2013}.}.   \AETtext{The absence of activity can be experimentally realised in two ways (Fig.~\ref{fig:scheme}). In the absence of ATP, A can still bind to the DNA-conjugated motors, and thus the average species C will still have the diffusion coefficient $D$. In the absence of motors, instead, the relevant diffusion coefficient would be $D_A$.} 

The overdamped equation of motion for the velocity field, relevant for slow flows encountered in biological and biomimetic systems, has the form
\begin{equation}
(\gamma -\eta \partial_{xx}) v =\partial_x [\zeta (c)]
\label{eq:fbalmine}
\end{equation}
where $\gamma$ is a friction coefficient that accounts for relative motion with respect to the surrounding medium, $\eta$ is the viscosity of the active layer and $\zeta(c)$ is a stress that depends on the regulator concentration. This stress may have a passive component arising from an equation of state for $c$, but we take it to be primarily active with $\zeta(c)>0$ implying a contractile system and $\zeta(c)<0$ an extensile system.
For much of this paper, we will assume a form for $\zeta(c)$: $\zeta(c) = \alpha c$, where $\alpha$ is a constant. However, we checked that our results do not depend qualitatively on the form of $\zeta$ by also considering two alternate forms of the stress coupling: a quadratic form, $\zeta_c = \alpha c^2/c_0$, and a saturating form $\zeta(c) = \alpha c/(1 + c/c_0)$ (see appendix ~\ref{app:alt-sims} and Fig.~\ref{fig:alt-sims}). Eq. \eqref{eq:cons1mine} and \eqref{eq:fbalmine} fully characterise the effectively one-dimensional RDAM system.

To facilitate numerical calculations, we now non-dimensionalize \eqref{eq:cons1mine} and \eqref{eq:fbalmine} by rescaling $x\to x/\ell$, $t\to t/\tau$ and $c\to c/c_0$ where $\ell=\sqrt{\eta/\gamma}$ is the hydrodynamic correlation length which is proportional to the thickness of the active layer and $\tau=1/r$ is the zero-wavenumber inverse relaxation (growth) rate of concentration fluctuations. The rescaled equations of motion are
\begin{align}
&\partial_t c = \frac{1}{\mathrm{Da}}\partial_{xx} c - \partial_x(vc) + c(1 - c) \label{eq:cnd}\\
&(1-\partial_{xx}) v  = \Omega(c) \partial_x c \label{eq:vnd}.
\end{align}
where $c$, $x$, $t$ and $v$ are now non-dimensional, and $\text{Da}=r\eta/D\gamma$ -- the second Damk\"ohler number -- and $\Omega (c)= \partial_{c}\zeta( c)/r\eta$ are non-dimensional parameters of the model. Da and $\Omega$ express ratios, respectively, of characteristic timescales of reaction to diffusion, and of active transport to reaction. Thus, $\Omega$ can also be expressed as $\Omega =$ Pe/Da,  where Pe is the more familiar P\'eclet number expressing the ratio of active to diffusive time scales, given in our case by $\text{Pe}(c) = \partial_{ c}\zeta( c)/D\gamma$.

To understand the behavior of this model, we first consider the linear stability of a state with a homogeneous concentration implied by \eqref{eq:cnd} and \eqref{eq:vnd}, and then numerically analyze how activity modifies the propagation of a concentration front into a state with $c=0$.

\section{Linear Theory}\label{sec_linear_theory}

\begin{figure}
\includegraphics[width = 0.8\linewidth]{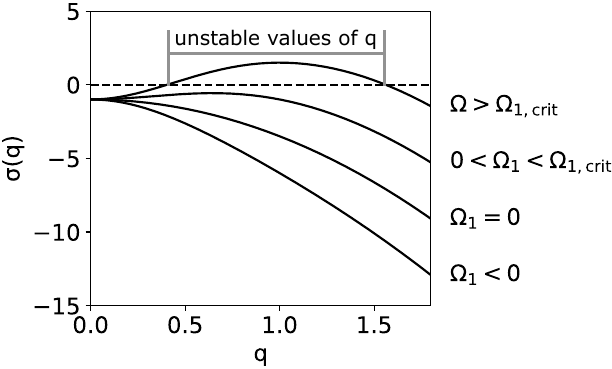}
\caption{Growth rate $\sigma$ of wavenumber $q$ at different values of the active stress parameter $\Omega_1 = \Omega|_{c=1}$, from linear stability analysis of \eqref{eq:cnd} and \eqref{eq:vnd} around $c = 1$ with a linear stress term $\zeta(c) = \alpha c$. The black dashed line is at $\sigma = 0$. For contractile stresses above a critical value of $\Omega_1$, given in \eqref{eq:omega_crit}, the reaction-diffusion active matter (RDAM) system is unstable to perturbations in a finite range of wavenumbers. Extensile stresses (negative $\Omega$) stabilize the system. In this figure, Da $= 0.4$.}
\label{fig:sigma-vs-q}
\end{figure}

\red{We start by examining the linear stability of a state at zero average concentration i.e., $c_{ss}=0$, with boundary conditions $c(\pm \infty) = 0$, $v(\pm\infty) = 0$.}
Linearizing \eqref{eq:cnd} and \eqref{eq:vnd} about $c_{ss}=0$ and Fourier transforming in space and time, we get 
\begin{equation}
\label{lin_css0}
    -i\sigma c_{q,\sigma}=\left(1-\frac{1}{\mathrm{Da}}q^2\right) c_{q,\sigma}
\end{equation}
where $\sigma$ and $q$ are the (non-dimensional) frequency and wavenumber of the perturbation respectively. The dispersion relation is $\sigma=i(1-q^2/\mathrm{Da})$ which implies that the amplitude of a small wavenumber fluctuation grows, destabilizing the $c_{ss}=0$ state. \red{Next, linearizing about $c_{ss}=1$ ($c(\pm \infty) = 1$, $v(\pm\infty) = 0$)} and defining $\delta c=c-1$ and $\Omega|_{c=1}\equiv\Omega_1$, we get
\begin{equation}
\label{disper}
    -i\sigma\delta c_{q,\sigma}=-\left(1+\frac{1}{\mathrm{Da}}q^2-\frac{\Omega_1 q^2}{1+q^2}\right)\delta c_{q,\sigma}.
\end{equation}
This implies that a state with $c_{ss}=1$ is stable in the long wavlength $q\to 0$ limit where $\sigma=-i$. Eq. \eqref{disper}, however, suggests an activity-induced \emph{nonhydrodynamic}, \emph{finite} wavenumber instability of the homogeneous state for wavenumbers in the range
\begin{widetext}
\begin{equation}
    |q|\in\left\{\frac{\sqrt{-[1+\mathrm{Da}(1-\Omega_1)]-\sqrt{[1+\mathrm{Da}(1-\Omega_1)]^2-4\mathrm{Da}}}}{\sqrt{2}}, \frac{\sqrt{-[1+\mathrm{Da}(1-\Omega_1)]+\sqrt{[1+\mathrm{Da}(1-\Omega_1)]^2-4\mathrm{Da}}}}{\sqrt{2}}\right\}
\end{equation}
\end{widetext}
when 
\begin{equation}
\Omega_1>\frac{1+2\sqrt{\mathrm{Da}}+\mathrm{Da}}{\mathrm{Da}}=\Omega_c
\label{eq:omega_crit}
\end{equation}
(or $\text{Pe}_1 > (1 + \sqrt{\text{Da}})^2$, where $\text{Pe}_1\equiv \text{Pe}|_{c=1}$), as sketched in Fig.~\ref{fig:sigma-vs-q}. Since this requires $\Omega_1>0$ it can only be achieved in \emph{contractile} systems. The maximum growth rate is at the wavenumber $q_c=\sqrt{-1+\sqrt{\Omega_1\mathrm{Da}}} = \sqrt{-1+\sqrt{\mathrm{Pe}_1}}$. \AMrev{The instability starts at finite wavenumber in our model, in contrast to the zero wavenumber instability predicted in \cite{Bois2011}, due to the reaction term which damps concentration fluctuations at zero wavenumber.} \footnote{A note of caution: a prediction of a finite wavenumber instability within a hydrodynamic theory should not be taken to imply that all systems with the same symmetry, and the same sign of activity, will display this instability, in contrast to prediction for a zero-wavenumber instability. Whether the growth rate turns positive at a non-zero wavenumber depends not only on the lowest order in wavenumber terms retained within the hydrodynamic theory but terms at all orders in wavenumbers that are not. Such terms encode system-specific details and can eliminate a finite wavenumber instability even when a theory retaining only the lowest order in wavenumber terms suggests its presence.}

\red{We now consider the case where a front moving at a constant speed $k$ interpolates between these two states, such that $c(-\infty) = 1$ and $c(\infty)=0$, and $v(\pm \infty) = 0$.} The steady-state front must then be described by a function $c(\xi)$, where $\xi=x-kt$ which, in the absence of flow-advection due to activity, solves the equation
\begin{align}
\label{KPP}
0 = \frac{1}{\mathrm{Da}}\partial_{\xi\xi} c + k\partial_\xi c + c(1-c).
\end{align}
The speed of the $c=1$ front in the absence of activity is expected to be, in dimensional form, $k=2(Dr)^{1/2}$, using a marginal stability argument originally due to  \cite{Dee}. Contractile or extensile activity qualitatively modifies the KPP concentration dynamics via the advection by the hydrodynamic velocity. Formally solving \eqref{eq:vnd} in terms of the concentration field leads to an integro-differential equation for the concentration which is not easy to deal with analytically. We will therefore examine effect of advection due to active flow on the front propagation numerically.
However, we can first gain some qualitative insight by assuming that $v\gg \partial_{xx} v$ in \eqref{eq:vnd}, which is generically the case when $\ell\to 0$, and replace the velocity field in \eqref{eq:cnd} with $v=\Omega(c) \partial_x c$ to obtain
\begin{align}
\label{KPPex}
0 = \left(\frac{1}{\mathrm{Da}} -c\Omega\right)\partial_{\xi\xi} c + \left[k-(c\Omega'+\Omega)\partial_\xi c\right]\partial_\xi c + c(1-c)
\end{align}
where the prime denotes differentiation with respect to $c$. Therefore, the coupling to an active species leads to two distinct modifications of the usual KPP equation: i. it reduces the (non-dimensional) diffusivity at non-zero concentrations via the term $c\Omega$ and ii. it leads to a new nonlinear term $(c\Omega'+\Omega)\partial_\xi c$ in the equation of motion. The former implies that a $c=1$ concentration profile behind the front cannot be homogeneous for large values of $\Omega$, as seen in the linear stability analysis.
Note, however, that linearizing \eqref{KPPex} suggests an incorrect value for the threshold of the instability of the $c_{ss}=1$ state discussed earlier, the range of unstable wavenumbers, and the wavenumber at which the growth rate of the instability is maximum. It predicts an instability for $\Omega_1=1/\mathrm{Da}$ and, since there is no stabilizing mechanism at large wavenumbers, no fastest growing mode.  The reason for this is that we assumed the hydrodynamic correlation length $\ell$ to be $0$ which amounts to replacing the factor $1/(1+q^2)$ in the last term of the R.H.S. in \eqref{disper} by $1$. If we expand $1/(1+q^2)$ in a binomial series for $q\to 0$, we see that this approximation ignores even \emph{linear} terms at higher wavenumbers. Therefore, this implies that \eqref{KPPex} becomes qualitatively different from the full dynamics when $\Omega\gtrsim 1/\mathrm{Da}$ and cannot be used to obtain qualitative information about it beyond this point.

If instead of setting $\ell\to 0$ and ignoring the $\partial_{xx} v$ term in \eqref{eq:vnd}, we formally invert the differential operator $(1-\partial_{xx})^{-1}\equiv\sum_{n=0}^\infty\partial_x^{2n}$, we find that the non-local character of the full velocity field leads to linear and nonlinear higher order in gradient terms in the concentration equation which should lead to wavenumber selection behind the front when the homogeneous state is unstable.

Finally, even in the regime where \eqref{KPPex} is predictive, the nonlinear modification of diffusivity and speed implies that results regarding the speed of the front derived using marginal stability analysis for KPP equations cannot be carried over to this case  \cite{Dee, VanSaarloos1, VanSaarloos2}. In particular, if \eqref{KPPex} is linearized around $c=0$, it simply reduces to the usual KPP equation \eqref{KPP} which would predict that the front speed would be unaffected by activity $\Omega$. This turns out to be incorrect, as we explicitly demonstrate in the next section by numerically simulating \eqref{eq:cnd} and \eqref{eq:vnd}.

\begin{figure*}
\includegraphics[width = \linewidth]{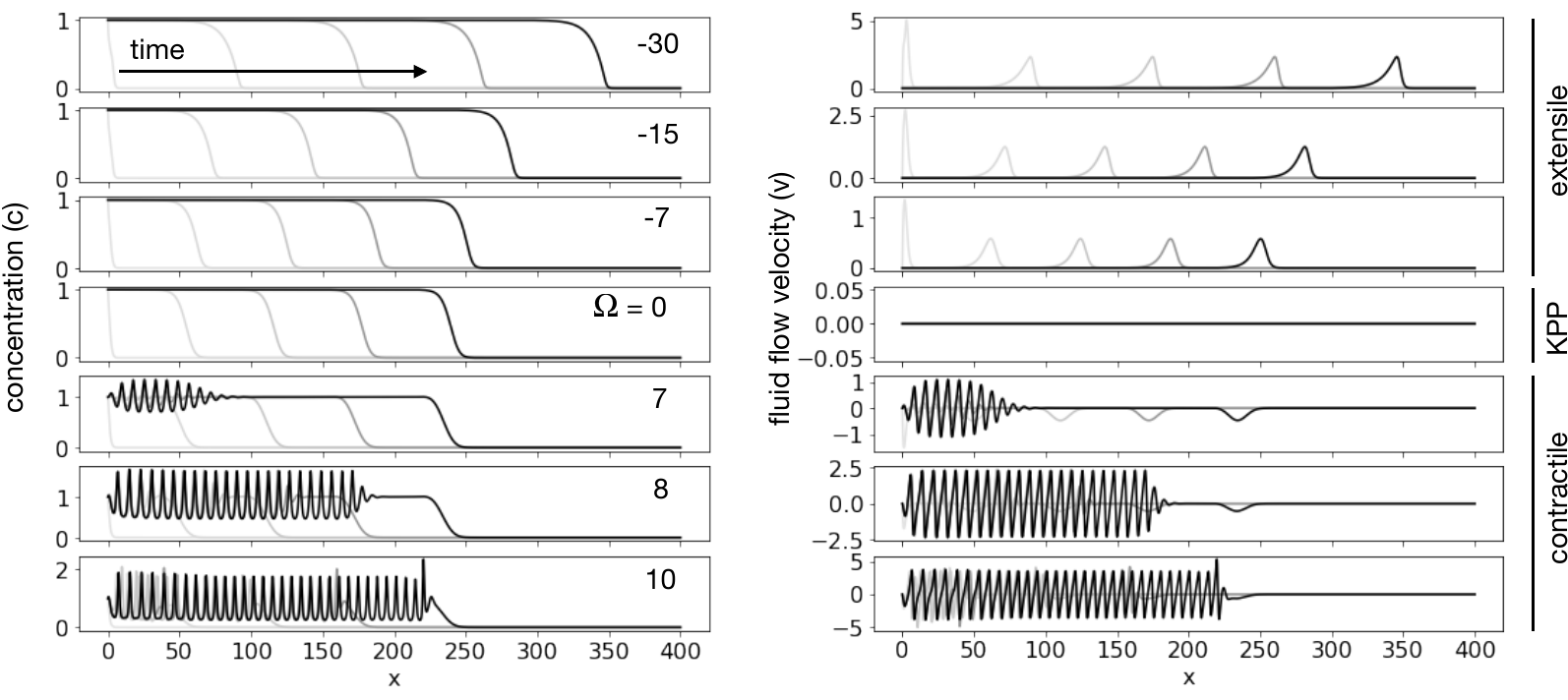}
\caption{Representative simulations of the RDAM model (\eqref{eq:cnd}-\eqref{eq:vnd}) studied in this article. This model of mechanochemical coupling may represent, for example, a simplified 1D version of an experiment in a thin film where a reacting and diffusing chemical species A interacts with stress-generating molecules (for instance, molecular motors walking on filamentous proteins) such that the magnitude of the stress increases with increasing concentration of A. Positive values of $\Omega$ model contractile stresses, while negative values model extensile stresses. Similar to Ref.~\citenum{Bois2011}, we find that a periodic instability develops behind the front above a critical value of contractile stresses. Moreover, we note several other interesting features which we explore in this paper. First, there is an asymmetric effect of stresses on the speed of the traveling front, which is increased by extensile stresses but unaffected by contractile stresses. Second, the instability is modified by the boundary condition imposed by the front in three ways: the critical value of $\Omega$ is shifted to a higher value compared to an infinite system; there is a gap between the front and the instability which closes with increasing $\Omega$; and the instability selects a single wavelength from the spectrum of unstable wavelengths in the ideal system. Simulation parameters for all simulations in this figure: $L = 400$, $T = 100$, $dx = 0.01$, $dt = 0.1$, $\mathrm{Da} = 0.4$.}
\label{fig:sims}
\end{figure*}

\section{Numerical results}

To explore the propagation of a concentration front in the RDAM system we numerically simulated \eqref{eq:cnd} and \eqref{eq:vnd}. In Fig.~\ref{fig:sims} we show example trajectories of these simulations. In this section we consider only stress linear in $c$, i.e. $\zeta(c) = \alpha c$, so that $\Omega$ is independent of $c$ and we drop the subscript on $\Omega_1$.  Further examples are shown in the appendix Fig.~\ref{fig:alt-sims} with alternative forms of the stress term $\zeta(c)$ and the reaction term $f(c)$. We simulated the system using custom python code, which is available along with the analysis scripts used in this paper at \url{https://github.com/cdeljunco/rdam}. The coupled equations were integrated in real space using an implicit Euler scheme. Discretization parameters (grid size $dx$, time step $dt$) were chosen to be small enough that the simulations were numerically stable and results did not depend on them. Box length $L$ and total time $T$ were chosen to be large enough that the simulation results reached their asymptotic values. \red{Dirichlet boundary conditions were imposed on $c$ and $v$ with $c(0) = 1$, $c(L) = 0$ and $v(0) = v(L) = 0$, implying $\partial c_x(0) = \partial c_x(L) = 0$ due to equation~\eqref{eq:vnd}.}Da$ = \frac{r\eta}{D\gamma}$ was chosen to match experimentally realizable conditions with $r = \tau_A^{-1} = 10^{-2}-10^{-4}$~s$^{-1}$, the rate of the DNA/enzyme autocatalytic reaction and $D=D_A = 10^{-10}$~m$^2$s$^{-1}$, the diffusion coefficient of a 10-mer single-stranded DNA  \cite{zadorin2015}. For a thin film of thickness $h$, $\eta/\gamma\sim h^2$, yielding Da $= 1-10^{-2}$ for $h=100$~$\mu$m. We chose Da ranging from 0.2 to 0.8 in the following. Finally, the stress parameter $\Omega$ was varied in the range -30 to 12 to probe different regimes of contractile and extensile stresses. Specific parameter values are stated along with the results in what follows. 

Fig.~\ref{fig:sims} summarizes our central findings. It shows that with contractile stresses the front speed does not change significantly and, above a critical value of $\Omega$, an instability appears behind the front where the concentration is high, as predicted from the linear theory. In contrast, in the extensile case, the selected front speed $k^*$ increases as a function of $\Omega$ but no spatial instability is observed. We now examine these results in more detail.

\subsection{Front speed selection in the presence of activity \label{sec_num_results}} 

In this section, we examine how activity modifies the speeds of the concentration fronts. We defined the front location as the rightmost point where $c = 0.5$ (when the contractile instability sets in, there can be points behind the front with $c = 0.5$). We computed the location of the front from the simulations in Fig.~\ref{fig:sims} and found that it is a linear function of time after an initial transient resulting from the step function initial conditions that we used. This is because the speed of a KPP front and its shape are related; specifically, linearizing \eqref{KPP} about $c = 0$ and assuming that $c(x, t)$ is a traveling wave with the exponential form $c(x, t) \propto  \exp(-a(x - kt))$ at the leading edge, yields a dispersion relation which predicts that the front speed $k$ decreases with increasing $a$, saturating at $k = 2\mathrm{D r}^{-1/2}$ for $a \geq 1$~\cite{Murray2002}.  As a result, simulations begun from an infinitely steep front must accelerate to their asymptotic speed. We therefore calculated the front speed from a linear fit of the front position vs. time at late times after the speed had stopped changing. We verified our results by checking that we obtained the same front speed in simulations initialized with a shallow initial profile of the form $c(x, 0) = \tanh((x - x_0)/m)$ with $m$ chosen to be large enough that the front speed converges to its final value from above. Additionally, we checked that our choices of lattice spacing ($dx = 0.5$) and time step ($dt = 0.1$) were small enough that further reducing them had a negligible effect on the calculated front speed. 

We first consider the effect of extensile activity on the front speed. In our reduced model \eqref{KPPex}, if $\Omega$ is independent of $c$, the activity \emph{nonlinearly} enhances the $\partial_{\xi\xi}$ term and reduces the $\partial_\xi$ term compared to the standard KPP equation, making it quite different from the standard KPP equation \eqref{KPP}. It follows from a simple physical argument that restricts solutions to have non-negative values of $c$ that a KPP front can adopt any speed above a minimum allowed value~\cite{Murray2002}, but it has been shown using marginal stability analysis that in fact the front asymptotically picks out this minimum speed~\cite{Dee}. While a marginal stability argument is not expected to yield the front speed in this case, we hypothesize that the selected front speed in full simulations ($k^*$) will still be given by the minimal allowed value of $k$, $k_{min}$, for which $c$ is non-negative in simulations of \eqref{KPPex}.
We \AMrev{numerically integrated} \eqref{KPPex} for each value of $\Omega$ using Matlab's ode23s function for a range of values of $k$ (code online at \url{https://github.com/cdeljunco/rdam}). 
The minimum allowed \AMrev{value of the front speed,} \AMrev{$k_{min}$}, was identified as the smallest value of $k$ for which \AMrev{$c\geq0$ for all $\xi$}. The $k_{min}$ obtained \AMrev{via this procedure} is an increasing function of $\Omega$ (in contrast to the prediction from a naive linearization of \eqref{KPPex} about the $c=0$ state~\cite{Dee}). This observation can be heuristically accounted for by the following argument. For \eqref{KPP}, linearizing about $c = 0$ shows that, due to the constraint that $c\geq0$, the only possible trajectories have $k \geq 2(Dr)^{1/2}$~\cite{Murray2002} (which is also the speed of the front~\cite{Dee}).
Now, since $c$ must always be positive and $\partial_\xi c$ is always negative in permitted trajectories where the front moves from left to right, in \eqref{KPPex} a negative value of $\Omega$ (corresponding to extensile stresses) increases the diffusivity ($D_{\text{eff}}  = 1/\text{Da} - c\Omega$). At the same time, the part of $k_{\text{eff}}$ (which is the coefficient of $\partial_\xi c$ in \eqref{KPPex}) that depends on $\partial_\xi c$ is $- \Omega\partial_\xi c$ and therefore decreases.
%nonlinearly decreasing the coefficient of the gradient term
 Since $k_{\text{eff}}$ is constrained to be greater than $D_{\text{eff}}$, and since $D_{\text{eff}}$ increases and $- \Omega\partial_\xi c$ decreases as $\Omega$ becomes more negative (i.e., for increasing extensile stress), the value of $k$ must increase to compensate. Thus, the allowed values of front speed shift upwards. 

\begin{figure}
\includegraphics[width = 0.8\linewidth]{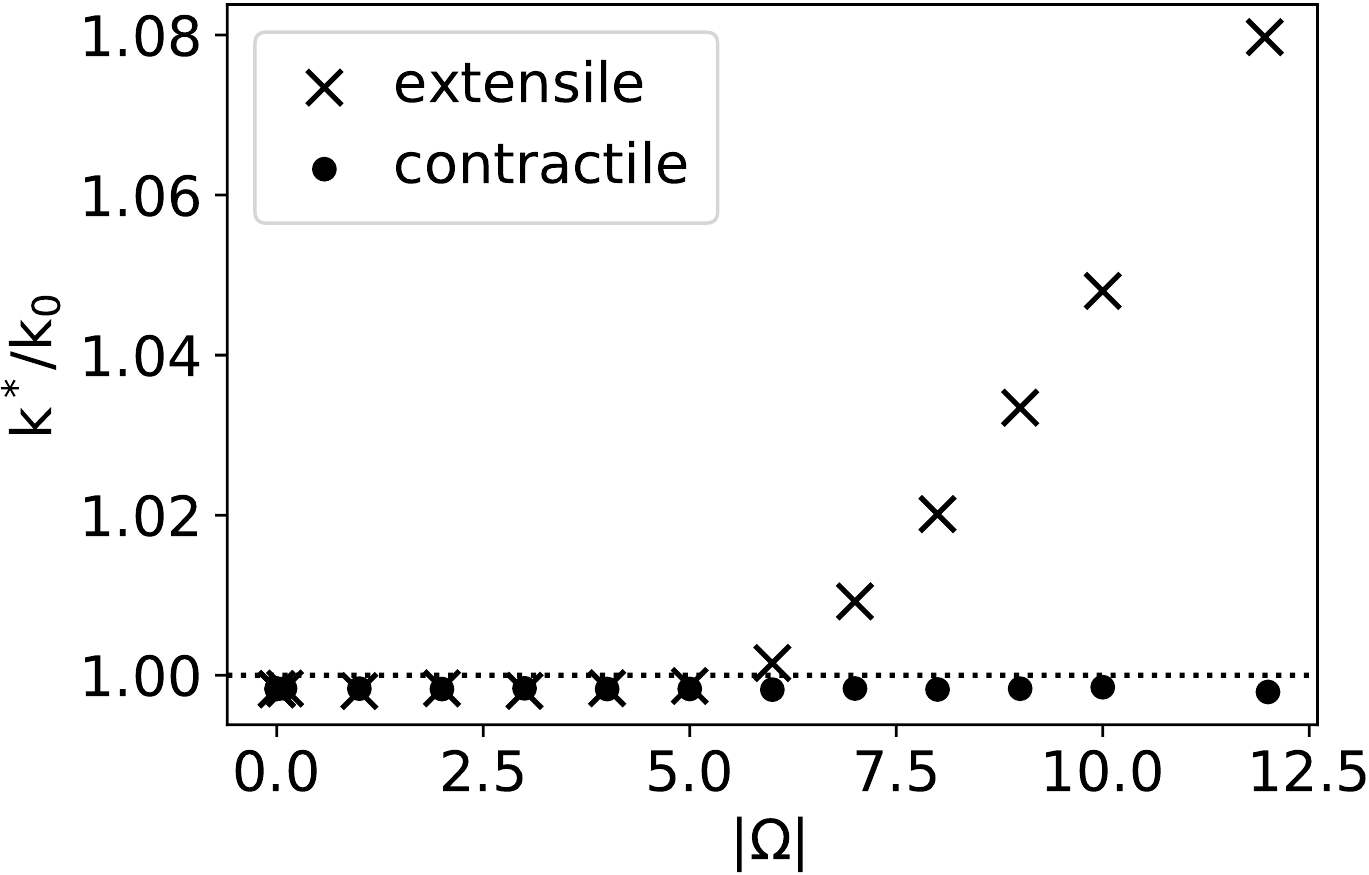}
\caption{The effect of extensile and contractile stresses on the selected front speed $k^*$. Extensile active stresses result in an increase in $k^*$ above a critical value $\Omega^*$ of $\Omega$, while contractile stresses have no effect. All y-axes are normalized by $k_0 = 2\mathrm{Da}^{-1/2}$, the front speed at $\Omega = 0$. The line at $k^* = k_0$ is a guide for the eye. Simulation parameters for these plots: $T = 900$, $dt = 0.05$, $dx = 0.1$, $L = 3000$ for $\Omega > -7$ and $L = 2800 + 58|\Omega|$ for $\Omega \leq -7$. $L$ was chosen such that the front would not reach the end of the box before $t = 900$.}
\label{fig:k-ext-con}
\end{figure}

\begin{figure}
\includegraphics[width = 0.8\linewidth]{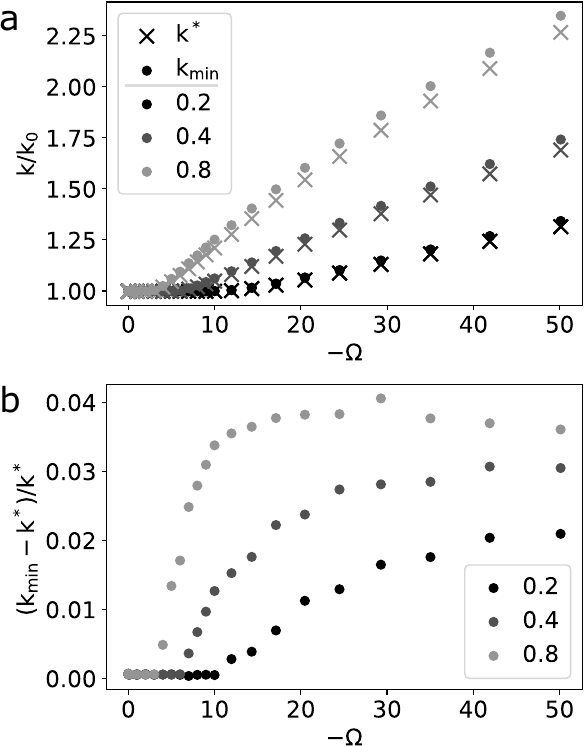}
\caption{The front selects the minimum allowed speed. (a) Comparing the front speed selected in simulations of the full model ($k^*$) to the minimum allowed front speed $k_{min}$ in a simplified model where we take $\ell \to 0$. Shades of grey represent different values of Da (0.2, 0.4, 0.8), x's are results from a simulation of the full system, and circles are results from the simplified model. The agreement is good over this range of parameters and improves with decreasing Da. (b) Relative difference between the data from full and simplified models in (a). Simulation parameters for these plots: $T = 900$, $dt = 0.05$, $dx = 0.1$. For Da $= 0.2$ and $0.4$, $L = 3000$ for $\Omega > -7$ and $L = 2800 + 58|\Omega|$ for $\Omega \leq -7$. For $Da = 0.8$, $L = 3000$ for $\Omega > -6$ and $L = 3000 + 75|\Omega|$ for $\Omega \leq -7$.}
\label{fig:front-speed}
\end{figure}

We now compare this result with the full numerical simulation to test our hypothesis. Fig.~\ref{fig:k-ext-con}a shows that the front speed begins to increase in extensile simulations for $\Omega \approx -5$ (for $\mathrm{Da}=0.4)$. In Fig.~\ref{fig:front-speed} we compare $k^*$ to $k_{min}$ for three values of Da and $0 < |\Omega| < 25$.  Fig.~\ref{fig:front-speed}b shows that $k_{min}$ is within 5\% of $k^*$ for the largest value of Da that we looked at ($\mathrm{Da} = 0.8$), and is within 1\% for the smallest value $\mathrm{Da} = 0.2$. This is a significant finding for two reasons: i. Eq. \eqref{KPPex} was obtained within an $\ell\to 0$ approximation and therefore ignored both linear and nonlinear terms of higher order in gradients. Despite this, the reduced model predicts the front speed reasonably accurately. ii. This demonstrates that even in the presence of activity, which makes the effective concentration equation non-local due to the non-local dependence of the velocity field on the concentration, the selected speed is still given by the minimum allowed speed as had been predicted by  \cite{Dee, VanSaarloos1} in the context of much simpler models. The selected speed in both the full simulation and from \eqref{KPPex} remains equal to the speed in the absence of activity up to a critical value $\AMrev{|\Omega^*|}$ and then starts to increase with \AMrev{$|\Omega|$} as shown in Fig.~\ref{fig:front-speed}a. Fig.~\ref{fig:front-speed}a also demonstrates that $\Omega^*$ decreases with increasing $\mathrm{Da}$. This feature may also be heuristically understood from \eqref{KPPex}: when $c\mathrm{Da}\AMrev{|\Omega|}\ll 1$, the effective diffusivity in \eqref{KPPex} is effectively controlled by $1/\mathrm{Da}$ and the speed is equal to the one in the absence of activity. However, as Da increases, $\AMrev{|\Omega|}\sim 1/\mathrm{Da}$ at smaller values of $\Omega$ implying that the front speed is controlled by activity for $\AMrev{|\Omega| > |\Omega^*|}$ that decreases with Da.

\AETtext{Comparing the results in Fig.~\ref{fig:front-speed} with the values of the experimental parameters summarized in section~\ref{sec_model} allows us to evaluate the magnitude of the increase in the front velocity expected in the experiment sketched in Fig~\ref{fig:scheme}. Taking the thin film approximation $\eta/\gamma\sim w_d^2$, we have $Da = w_d^2\frac{1}{D_A\tau_A}\approx 2\times(10^{-3}-10^{-1})$ for a reasonable value of $w_d= 40~\mu$m. Kinesin/microtubule active fluids generate flows with velocities up to\cite{bate2019collective} $v_a=600~\mu$m/s associated with typical lengths $l_a = 100~\mu$m, yielding $Pe= v_al_a/D_A \approx 6$ and thus $\Omega = Pe/Da \approx 6\times (10^3-10)$. In these conditions, from Fig.~\ref{fig:front-speed}, we expect a change in the front velocity of about 20\% in the presence of activity, which could be reasonably measured experimentally.}

We now consider the effect of contractile active stress on the front propagation. As discussed earlier, \eqref{KPPex} is only expected to be predictive for \AMrev{$\Omega<1/\mathrm{Da}$}. We find that, in this regime \AMrev{(i.e., when \eqref{KPPex} does not imply an instability of the high concentration state)}, \eqref{KPPex} implies that the minimum allowed value of $k$ is essentially the same as the one for $\Omega=0$. \AMrev{This is heuristically understandable: even for the extensile system, the front speed was only controlled by activity, and increased with it, when $|\Omega|\gtrsim 1/\mathrm{Da}$. Therefore, since \eqref{KPPex} is only valid for $\Omega<1/\mathrm{Da}$, we don't expect the front speed to decrease with contractility at least when the high concentration front is not itself unstable.} Our results from the full numerical simulations confirm that the front speed is not modified when the active stress is contractile, as shown in Fig.~\ref{fig:k-ext-con}. Surprisingly, we find that the front speed remains independent of $\Omega$ even at larger values of $\Omega$ when the concentration and velocity profiles behind the front develop a periodic pattern, which we discuss in the next section.

\subsection{Patterns behind the front with contractile stresses\label{sec:instab}}

\begin{figure}[h!]
\includegraphics[width = 0.8\linewidth]{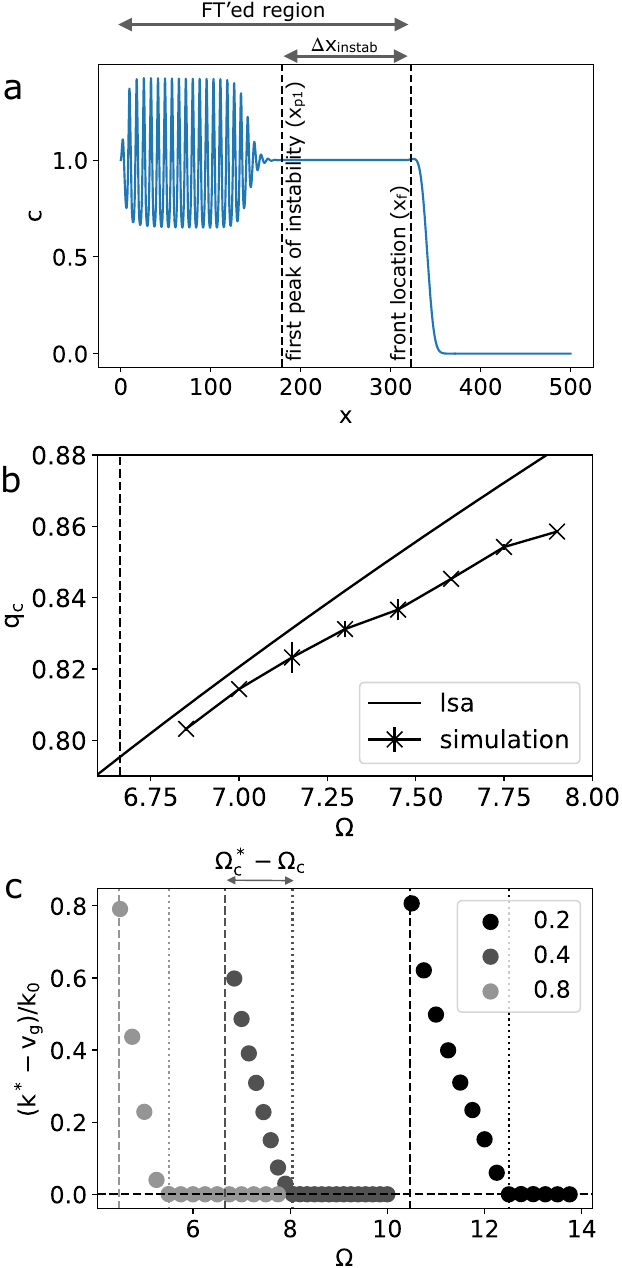}
\caption{Wavenumber and group velocity of the periodic pattern behind the front. (a) To measure the group velocity $v_g$ and the fastest growing wavenumber $q_c$, we defined the location of the front $x_f$ as the rightmost peak of the concentration, and the location of first peak of the instability $x_{p1}$ as the location of the second rightmost peak. (b) $q_c$ as a function of $\Omega$, defined as the mean value over the final third of a long simulation of the wavenumber with maximum amplitude in the Fourier transform of the concentration between $x = 0$ and $x = x_f$. The vertical dashed line is located at $\Omega_c$. Parameters for these simulations were: Da=0.4, $dt = 0.05$, $dx = 0.05$, $L = 6300$, $T = 1800$.  (c) Front speed $k^*$ minus the group velocity of the periodic pattern $v_g$ as a function of $\Omega$ at three values of Da (colors). Each simulation data point is obtained from the slope of the distance from the front to the first peak of the instability, $\Delta x_{instab} = x_f - x_{p1}$, as a function of time, which is a linear function at long times. The vertical dashed lines are located at $\Omega_c$, and the dotted lines at $\Omega_c^*$, showing decreasing $\Omega_c^* - \Omega_c$ with increasing Da. Parameters for these simulations were: $dt = 0.05$, $dx = 0.05$, $L = 500$, $T = 150$.}
\label{fig:front-instab}
\end{figure}

We now consider the dynamics at large $\Omega$ when the homogeneous profile behind the front is destabilized and a periodic pattern is formed. We measure $\Omega_c$ -- the minimum value of $\Omega$ for which a periodic pattern emerges -- and $q_c$ -- the wavenumber of the periodic pattern -- from the simulation of the full model. We checked that in simulations with periodic boundary conditions $\Omega_c$ and $q_c$ are consistent with linear stability analysis predictions (see Appendix~\ref{app:lsa} for methods and results).  

In simulations with a front, there is a small peak in the concentration just behind the front, which we identified using a standard peak-finding algorithm as the peak with the largest value of $x$. We call the location of this peak $x_f$ as it tracks the location of the front. We define the leading edge of the instability $x_{p1}$ to be the location of second leftmost peak. We then Fourier transformed the region from $x = 0$ to $x = x_{f}$, as shown in Fig.~\ref{fig:front-instab}a. Because $x_{p1}$ is a function of time, the region of the Fourier transform changes over time and so do the possible wavenumbers of the instability.  We therefore approximated the height of the dominant wavenumber at long times ($q_c$) by the amplitude $h_{max}$ of the wavenumber with maximum amplitude ($q_1$) at any point in time. We plotted $\log(h_{max})$ vs. $t$ and fit the linear region of the curves to extract the growth rate $\sigma(q_1; \Omega)$. We then fit $\Omega$ vs. $\sigma(q_1)$ to a line and extrapolated $\Omega_c$ as the value of $\Omega$ where $\sigma(q_1) = 0$. 

Surprisingly, we find from this analysis that $\Omega_c\approx (1+2\sqrt{\mathrm{Da}}+\mathrm{Da})/\mathrm{Da}$ as predicted by the linear stability analysis of a homogeneous $c_{ss}=1$ state. Next, examining time-evolution of the dominant wavenumber $q_1$, we find that it relaxes from below to a constant mean value $q_c$ after $t \approx 1000$. In Fig.~\ref{fig:front-instab}b, we plot the mean $\pm$ 1 standard deviation of $q_1$ from $t = 1200 - 1800$ (i.e., $q_c$ $\pm$ 1 standard deviation), measured at intervals of $\Delta t_{sample} = 2$. The value of $q_c$ is within 1\% of the value predicted by the linear stability analysis when $\Omega$ is just above $\Omega_c$, and diverges from it as $\Omega_c$ increases. 

Finally, we measured the lag of the leading edge of the instability behind the front (i.e. the front speed $k^*$ minus the periodic pattern's group velocity $v_g$).  After a brief initial transient, the distance $\Delta x_{instab} = x_f - x_{p1}$ is a linear function of time. We calculated $k^* - v_g$ as the slope of the linear fit to $t$ vs $\Delta x_{instab}$. As shown in Fig.~\ref{fig:front-instab}c, $k^* - v_g$ goes from a positive value just beyond $\Omega_c$ to $0$. That is, the pattern lags behind the travelling front till a second value $\Omega_c^*$. Before this value, the front ``outruns'' the periodic pattern implying that near the travelling edge, a homogeneous region emerges before the patterned state sets in (see supplemental movie). Beyond $\Omega_c^*$, the pattern ``catches up'' with the front and for all higher values, moves with the front. The value of $\Omega_c^*-\Omega_c$ decreases with increasing Da (which depends on the ratio of the reaction coefficient and the bare diffusivity). This implies that in highly diffusive species, the patterns formed due to the contractile instability lag arbitrarily far behind the advancing front and will not be observed in experiments that track the front. Experimentally, microtubule active matter also displays a contractile behavior and could thus be used to test these predictions, either with multiheaded kinesins\cite{torisawa_spontaneous_2016}, clustered kinesins\cite{senoussi2021_RDAMbis, Bezia2021} or dyneins\cite{foster2015active}.

\section{Conclusion}

In this paper we examined the effect of active flows on the propagation of a reacting and diffusing activity regulator species. We demonstrated that neither contractile nor extensile active flows inhibit the propagation of the activity regulator. We further showed that extensile activity (corresponding to negative $\Omega = \mathrm{Pe}/\mathrm{Da}$ in our model) increases the propagation speed of the activity regulator relative to $k_0$, the speed in the absence of activity, when the magnitude of $\Omega$ exceeds a critical value $\Omega^*$, while contractile activity does not modify the front speed. On the other hand, contractile active flows above a critical value $\Omega_c$ destabilize the homogeneous state behind the front leading to a periodic pattern. For moderate Da and activity, however, a homogeneous state prevails near the front edge as the speed of the pattern lags behind the front speed.

RDAM systems may describe a wide range of {\it in vivo} biological systems. For instance, they may be relevant for the dynamics of activity regulators  \cite{Bois2011} in the cellular cortex or may describe the dynamics of cell monolayers  \cite{Oreffo}. They may also be synthesized {\it in vitro}. Indeed, we motivated our 1-dimensional, 1-variable model from the state-of-the art {\it in vitro} experiments involving a DNA/enzyme autocatalytic RD front coupled with a DNA-responsive kinesin/microtubule active fluid. We believe that the qualitative features predicted by the theory presented here can be observed in {\it in vitro} RDAM systems such as the one suggested here. Taken together, our results may prove useful in interpreting pattern formation in RDAM systems {\it in vivo} and {\it in vitro} that are on the experimental horizon.

\begin{acknowledgements}

We thank Rapha\"el Voituriez, Suri Vaikuntanathan, Jean-Christophe Galas, Anis Senoussi and Yuliia Vyborna for helpful discussions. This work has been funded by a Natural Sciences and Engineering Research Council of Canada Postgraduate Scholarship - Doctoral (NSERC PGS-D, C.d.J.); by the University of Chicago Materials Research Science and Engineering Center (MRSEC,  C.d.J.), which is funded by the National Science Foundation under Award No. DMR-1420709; and by the European Research Council (ERC) under the European's Union Horizon 2020 programme (grant No 770940, A.E.-T.).

\end{acknowledgements}

%\bibliography{main.bib}


%==========END BIBLIOGRAPHY===============

\begin{appendix}

\section{Demonstration that RDAM model is applicable both for aligned and isotropic systems}
In the main text, we described a simplified RDAM model in which we did not take orientational ordering of the active units into account. However, active matter systems, such as the kinesin-microtubule one described in the main text, are often aligned  \cite{senoussi2019}. In this appendix we show that the RDAM model is still applicable in this case as long as the speed of the KPP front is fast compared to the relaxation of the angular fluctuations in the aligned state. 

We consider an aligned system (for instance, composed of microtubules or polarised cells) with the degree of alignment denoted by 
\begin{equation}
\label{apol}
    {\bsf Q}=\frac{S}{2}\begin{pmatrix}1 && 0\\0 && -1\end{pmatrix}
\end{equation}
where $S$ is the degree of order with $S=0$ denoting an isotropic organisation and $S\neq 0$ denotes an aligned state. With this, the equation for the concentration field is modified to
\begin{equation}
\label{conceq}
\partial_t c = \bar{D}\nabla^2 c+\zeta_c \nabla\nabla:({\bsf Q}c) - \nabla\cdot({\bf v} c) + f(c)
\end{equation}
where $\bar{D}$ is the bare diffusivity and $f(c)$ is a generic reaction term. The ``curvature current'' $\propto \zeta_c$ is a standard active concentration current in active, aligned systems  \cite{RMP}.
The equation of the velocity field becomes
\begin{equation}
\label{veleq}
    (\eta\nabla^2-\gamma){\bf v}=-\nabla\Pi-\nabla \bar{\zeta}_c(c)-\nabla\cdot(\bar{\zeta}(c){\bsf Q})
\end{equation}
where $\Pi$ is the equilibrium pressure stemming from an equation of state for $c$, $\bar{\zeta}_c$ is the isotropic active stress and $\bar{\zeta}$ is the active apolar stress that arises in aligned systems  \cite{Aditi1, RMP}.
To complete the description of the model, we also need to specify the fluctuations of the order parameter ${\bsf Q}$. However, we assume that the propagation of the chemical species happens at a time-scale that is much faster than the alignment dynamics and on that timescale, the order-parameter ${\bsf Q}$ can be assumed to be frozen at a constant value. Therefore, we replace ${\bsf Q}$ by its value given by \eqref{apol} with $S=0$ for an isotropic state and $S=1$ for an aligned state. Further assuming that only variations of $c$ and ${\bf v}$ are along $\hat{x}$, we obtain the equations of motion described in the main text with $v\equiv v_x$, $D=\bar{D}+\zeta_c$ when $S=1$ (i.e. an aligned state) and $D=\bar{D}$ when $S=0$, $\zeta(c)=\bar{\zeta}(c)+\bar{\zeta}_c(c)-\Pi$ in the aligned state and $\zeta(c)=\bar{\zeta}_c(c)-\Pi$ in the isotropic state. 

\section{Alternative stress and reaction terms}\label{app:alt-sims}

\begin{figure*}
\includegraphics[width = \linewidth]{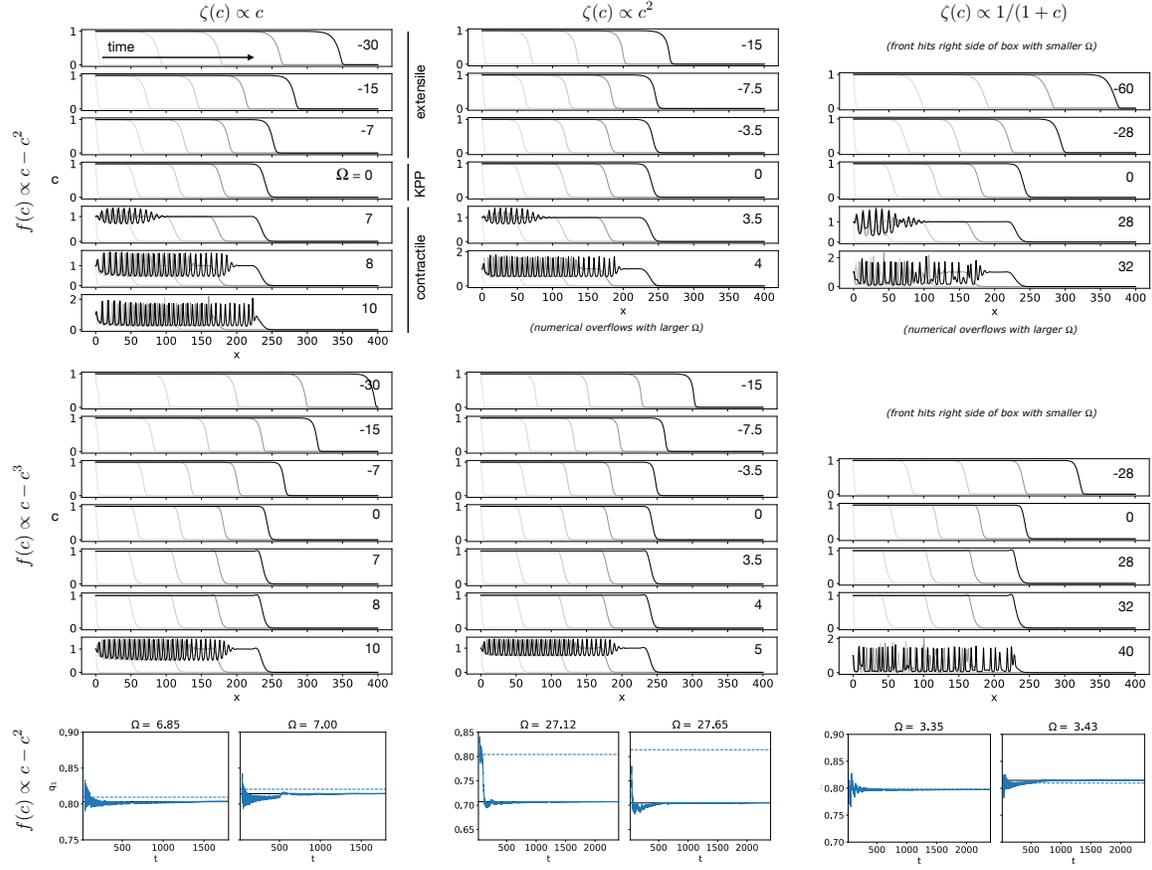}
\caption{Simulation results with alternative choices of stress and reaction terms. (Top two rows) Representative simulations of the concentration as a function of time in the reaction diffusion active matter (RDAM) model (\eqref{eq:cnd}-\eqref{eq:vnd}), with three alternative forms of the stress term $\zeta(c)$ and two forms of the reaction term $f(c)$. The upper left panel is the same as in Fig.~\ref{fig:sims}. Parameters: Da$ = 0.4$, $L = 400$, $T = 100$ for all simulations; $dx = 0.05$ and $dt = 0.1$ for simulations with $\zeta(c) \propto c$ and $\zeta(c) \propto c^2$; $dx = 0.01$ and $dt = 0.05$ for simulations with $\zeta(c) \propto c/(1+c)$. (Bottom row) Wavenumber with maximum amplitude in the Fourier transform of the concentration behind the front when $\Omega > \Omega_c$, as a function of time, computed as described in section~\ref{sec:instab}. Parameters: Da$ = 0.4$ for all simulations; $T = 1800$, $L = 6300$, $dx = 0.05$, and $dt = 0.05$ for simulations with $\zeta(c) \propto c$;  $T = 2400$, $L = 8400$, $dx = 0.05$, and $dt = 0.05$ for $\zeta(c) \propto c^2$; $T = 2400$, $L = 8400$, $dx = 0.01$, and $dt = 0.05$ for $\zeta(c) \propto c/(1+c)$.} 
\label{fig:alt-sims}
\end{figure*}

Fig.~\ref{fig:alt-sims} demonstrates the phenomena that we describe in this paper are also present with several alternative choices for the stress and reaction terms $\zeta(c)$ and $f(c)$. Specifically, for two reaction terms and three stress terms, we observe an increased front speed with extensile stresses and an instability behind the front that picks out a single wavelength at long times. 
Linear stability analysis predicts that instabilities arise when 
\begin{equation}
\Omega_1>(1+2\sqrt{\mu \mathrm{Da}}+\mathrm{Da})/(\beta\mathrm{Da}),
\label{eq:omega_c_alt}
\end{equation}
where $\mu$ and $\beta$ are constants that depend on the choice of $\zeta(c)$ and $f(c)$ with values given in Table~\ref{tab:constants}. Predicted values of $\Omega_{c}$ for each value of $\mu$ and $\beta$ are given in Table~\ref{tab:omega_c}. Fig.~\ref{fig:alt-sims}(a) shows that the location of the instability is shifted in agreement with the linear stability prediction for all choices of $f(c)$ and $\zeta(c)$ considered here.
While it is not clear from the short simulation in Fig.~\ref{fig:alt-sims}a that the model with saturating stress ($\zeta(c) \propto c/(1+c)$) picks out a single wavelength, the longer simulations in Fig.~\ref{fig:alt-sims}b show that at long times $q_1$ converges even in this case.

\begin{table}[h!]
\begin{tabular}{r|c|c|c}
$\zeta(c)/\alpha \to$ & $c$ & $c^2$ & $c/(1+c)$ \\ \hline
$\beta \to$ & 1 & 2 & 1/4 \\ 
\end{tabular}
\begin{tabular}{r|c|c}
$f(c)/r \to$ & $c-c^2$ & $c-c^3$ \\ \hline
$\mu \to$ & 1 & 2  \\ 
\end{tabular}
\caption{Values of the constants $\mu$ and $\beta$ in~\eqref{eq:omega_c_alt} for each choice of $\zeta(c)$ and $f(c)$ shown in Fig.~\ref{fig:alt-sims}.}
\label{tab:constants}
\end{table}

\begin{table}[h!]
\begin{tabular}{ c | c | c | c }
$\Omega_{1, c}$ & $\zeta(c) \propto c$ & $\zeta(c) \propto c^2$ & $\zeta(c) \propto c/(1+c)$  \\ \hline
  		$f(c) \propto c-c^2$	   & 6.66 & 3.33 & 26.65 \\ \hline
  		$f(c) \propto  c-c^3$	   & 8.97 & 4.49 & 35.89 \\ 
\end{tabular}
\caption{Critical values of $\Omega$ in non-dimensional units, when Da = 0.4, for each choice of $\zeta(c)$ and $f(c)$ featured in Fig.~\ref{fig:alt-sims} as predicted from a linear stability analysis.}
\label{tab:omega_c}
\end{table}

\section{Confirming linear stability analysis results in simulations}\label{app:lsa}

\begin{figure}
\includegraphics[width = 0.8\linewidth]{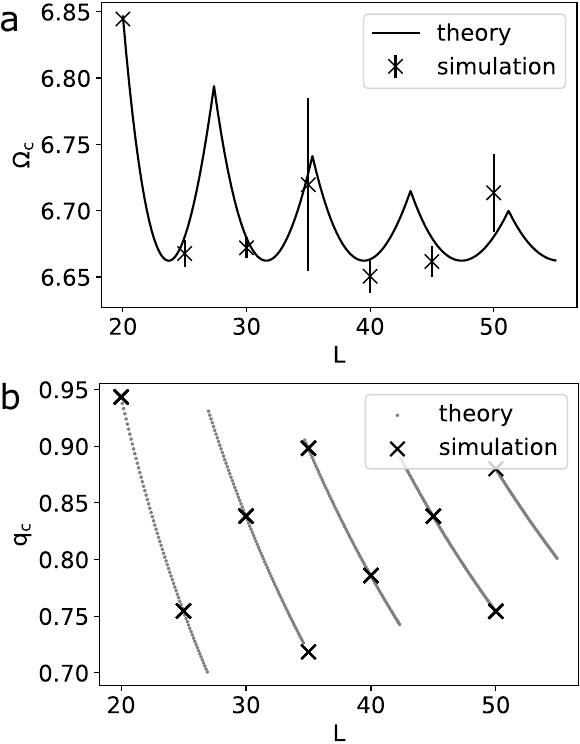}
\caption{Checking linear stability analysis predictions in simulations. (a) The critical value of the stress for instabilities to arise as a function of the box length. Error bars are $\pm 1$ standard deviation about the mean from 10 simulations with different random initial conditions. (b) The most unstable wavenumber $q_c$ at $\Omega  = 6.95$ as a function of the box length. All of the points from 10 simulations with different random initial conditions are plotted. Simulation parameters for all data in this figure are $dx = 0.01$, $dt = 0.1$, $T = 100$, $\mathrm{Da} = 0.4$.}
\label{fig:lsa}
\end{figure}

In this section we compare our linear analysis in section~\ref{sec_linear_theory} to simulations with periodic boundary conditions. Beginning from a system with a high concentration plus a random initial perturbation, $c(x) = 1 + 0.001\nu(x)$ where $\langle \nu(x) \rangle = 0, \langle \nu(x)\nu(x') \rangle = \delta(x-x')$, we simulated a system with parameters $\mathrm{Da} = 0.4$, $dx = 0.01$, $dt = 0.1$, $T = 100$, and a range of box sizes.  Denoting the Fourier transformed concentration as $\hat c$ and assuming that fluctuations in the concentration grow as $\hat c(q) \propto \exp(\sigma(q) t)$, we measured the growth rate of the fastest growing wavenumber, $\sigma(q_c)$, as the slope of the best fit line to $\log(c(q_c))$ vs. $t$. We then fit $\Omega$ vs. $\sigma(q_c)$ to a line and extrapolated $\Omega_c$ as the value of $\Omega$ where $\sigma(q_c) = 0$. We find that the critical values of $\Omega_c$ (Fig.~\ref{fig:lsa}a) and $q_c$ (Fig.~\ref{fig:lsa}b -- the wavenumber that dominates the spectrum of $c$ at long times), are in good agreement with our linear stability analysis calculations.

\end{appendix}

\end{document}